# Intense THz source based on BNA organic crystal pumped at Ti:Sapphire wavelength

MOSTAFA SHALABY,[1*+] CARLO VICARIO,[1*+] KARUNANITHI THIRUPUGALMANI,[2] SRINIVASAN BRAHADEESWARAN,[2] AND CHRISTOPH P. HAURI[1,3*]

[1] SwissFEL, Paul Scherrer Institute, 5232 Villigen-PSI, Switzerland
[2] Department of Physics, Bharathidasan Institute of Technology, Anna University, 620024 Tiruchirappalli, India
[3] EcolePolytechniqueFederale de Lausanne, 1015 Lausanne, Switzerland

[+] These two authors contributed equally to this work

*most.shalaby@gmail.com; carlo.vicario@psi.ch; christoph.hauri@psi.ch

**We report on high energy terahertz pulses by optical rectification (OR) in the organic crystal N-benzyl-2-methyl-4-nitroaniline (BNA) directly pumped by a conventional Ti:Sapphire (Ti:Sa) amplifier. The simple scheme provides an optical to terahertz conversion efficiency of 0.25% when pumped by a collimated laser pulses with duration of 50 fs and central wavelength of 800nm. The generated radiation spans frequencies between 0.2 and 3 THz. We measured the damage threshold as well as the dependency of the conversion efficiency on the pump fluence, pump wavelength, and pulse duration.**

High field terahertz (0.1-10 THz) generation is a topic of numerous ongoing researches due to its wide range of science applications [1-11]. Presently, optical rectification (OR) in nonlinear crystals is the preferred generation technique as it allows the realization of the most intense THz source. The figures of merit of a high-field THz source are the pump-to-THz conversion efficiency, THz spectrum, focusability and the applicable pump wavelength. The wide use of Ti:Sa laser technology in ultrafast THz spectroscopy has inspired the development of a THz emitter which can be pumped directly by those powerful lasers. Presently, for the most intense THz sources Ti:Sapphire needs to be converted to the short wavelength mid-infrared spectral range to provide efficient phase-matching [11-14].

In this letter, we report on THz generation in the organic crystal BNA which is pumped at the Ti:Sa 800 nm wavelength. The conversion efficiency and spectral density offered by BNA surpasses other organic crystals pumped at conventional Ti:Sa wavelength in the range of 0.1-3 THz by a factor 10-100. It thus represents a valuable option to the widely used Ti:Sa pumped Lithium Niobate (LN) crystal as the generation scheme is simpler and does not require pulse front tilting.

Among all nonlinear crystals, LN [10] and organic crystals (DAST, DSTMS, OH1) [11-15] provide most efficient OR. The typical conversion efficiency for LN pumped at 800 nm [10] and 1030 nm reaches 0.1-0.2 % that can thus be further increased by a factor of three by cryo-cooling the LN [15, 16]. However, the reported conversion efficiency is still debated as other research groups [15, 16] who repeated the experiment obtained conversion efficiency smaller by an order of magnitude than ref. [17, 18]. Due to the notorious focusing issues of LN, which are partly caused by the asymmetric beam quality and the complex geometry, the maximum reported THz field strength has not surpassed ≈1 MV/cm [10]. LN-based sources emit in the low-THz range (0.1-3 THz). Until now, this range has been difficult to access by organic crystals.

Organic crystals offer excellent phase-matching for OR in a simple pump beam configuration and superior THz focusability. The high 2-3% conversion efficiency is achieved for a pump wavelength between 1.2-1.5 μm [11, 13]. Such a scheme allows for extreme focusing peak fields reaching up to 83 MV/cm [11]. Yet, the main drawback of organic crystals is the required mid-IR laser pump wavelength for efficient phase-matching. In turn, this requires converting the Ti:Sa laser wavelength to longer wavelengths by using an optical parametric amplifier (OPA). In comparison, LN is efficient when pumped directly with 800 nm wavelengths. This fact has made LN the most commonly used THz generation approach so far.

There have been reports extending the use of organic crystals to the Ti:Sa wavelengths for intense THz sources. However, in these experiments, the maximum conversion efficiency was limited to $5\times10^{-5}$ [19, 20] and the emitted 1-5 THz radiation did not cover the scientifically interesting low-frequency THz range between 0.2-1 THz [21, 22]. These studies focused on adapting the mid-IR phase-matched crystals (DAST, DSTMS, TMS) to the 800 nm pump wavelength.

BNA is a highly nonlinear organic crystal well-suited for intense THz generation pumped at Ti:Sa wavelength [23-26]. It was developed by Hashimoto et al [23] as a derivative of MNA (an organic crystal with technical growth difficulties [27]). The most important criteria in THz generation from OR is the effective nonlinear optical coefficient $d_{eff}$ and the effective generation length over a certain spectrum. For BNA, $d_{eff}$ = 234±31 pm/V [25] which is comparable with the effective nonlinearity of other organic crystals (214 pm/V for DSTMS and 240 pm/V for OH1) [11]. The typical emission spectrum of LN [10] covers 0.1-2.5 THz for an 800 nm pump wavelength. At 800 nm pump wavelength, BNA has an effective generation length of 0.5-1 mm [24] and provides a THz spectrum similar to LN.

The BNA crystals were grown by isothermal solvent evaporation method at a temperature of 34 °C. We used 800 nm-centered Ti:Sa laser pulses (repetition rate 100 Hz, pulse duration 50 fs) to pump a 680 μm-thick BNA crystal. The experimental setup is shown in Fig. 1a, The generated THz field profile and spectrum are shown in Figs. 1b and 1bc respectively. The time trace shows a Mexican hat-like single cycle THz pulse. Such a pulse shape resembles the half-cycle THz pulse desired for applications such as magnetic domain flipping [28]. The spectrum has a cut-off at 3 THz and a center frequency of 1.3 THz. These characteristics are similar to that obtained from LN sources [10].

We studied the THz pulse energy and conversion efficiency as function of the pump fluence using a BNA crystal with a 2 mm clear aperture and a thickness of 680 μm. Although the THz energy is expected to scale quadratically with the pump energy, we experimentally obtained a nearly linear dependence (Fig. 2a). This is due to the nonlinear absorption of the pump and THz radiation inside the BNA crystal. At a fluence of 5.3 mJ/cm$^2$, we measured 0.54 μJ THz energy for a pump energy of 210 μJ using the normal spectrum from our laser (inset in Fig. 1a). This corresponds to a pump to THz conversion efficiency of 0.26 %. This is the highest achieved value in crystal-safe operation. As we increased the pump fluence further, the THz energy increased correspondingly but gradual damage of the crystal was observed over time. From these measurements we conclude that the damage threshold of BNA is more than 6 mJ/cm$^2$ for our laser parameters.

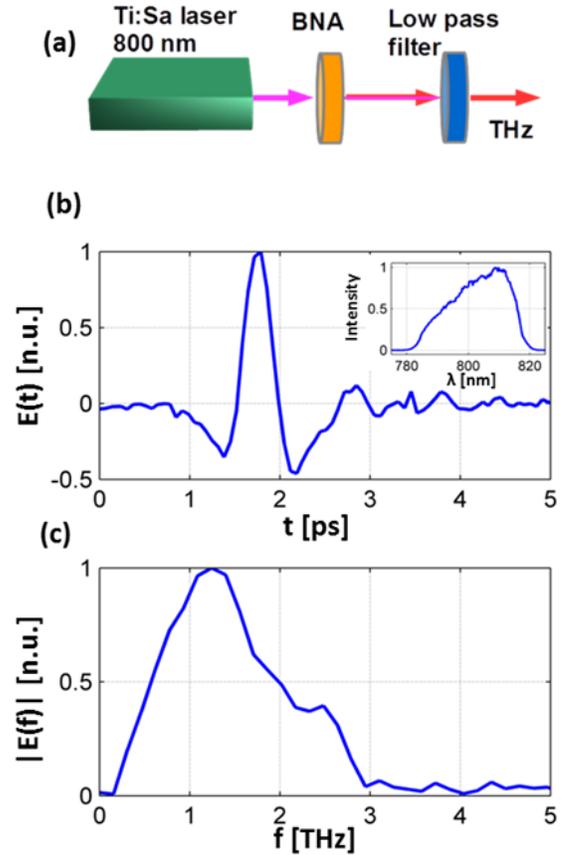

**Fig. 1**. Terahertz pulse characteristics. (a) the experimental setup. (b) the temporal evolution of the electric field and (c) the amplitude spectrum of the generated THz pulse. The spectrum of the laser pump pulse is shown as an inset in (b).

The high quantum conversion efficiency is accompanied by a small spectral red shift of the pump pulse spectrum. In Fig. 2b, we compare the transmitted pump spectrum after the crystal is set for phase-matched (red line) and non-phase matched conditions (blue line). In the latter case the pump to THz conversion is negligible. We observed a spectral red shift of the pump of nearly 1.6 nm corresponding to a conversion efficiency of 0.17 %, which compares well with the measured THz energy. A spectral red shift indicates the onset of cascaded OR, giving rise to enhanced conversion efficiency beyond the Manley Row limit. For BNA the calculated quantum conversion efficiency is close to 100%, which is below what has been observed in other organic crystals for a mid-infrared pump in the past [11-14].

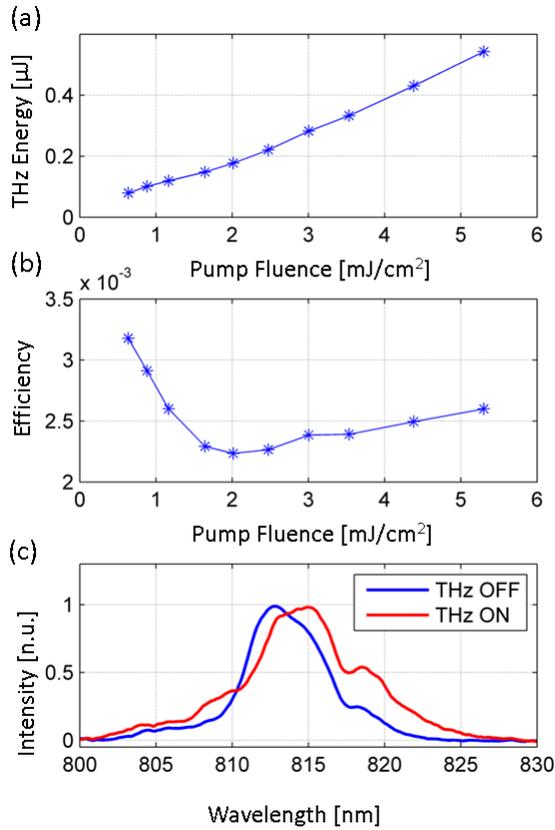

**Fig. 2.** Pump-THz conversion efficiency. (a) the generated THz energy and (b) efficiency versus the pump fluence. (c) the spectral shift of the 800 nm pump when the THz is applied.

Finally, we investigated the dependence of the generation efficiency on the pump wavelength and the pulse duration, respectively. The phase matching pattern for BNA around 800 nm is complex, as was previously shown by Miyamoto et al [24]. Moreover, the pump pulse duration used in our experiment is far from the best phase-matching. At a wavelength of 800 nm, the effective bandwidth over 0.5-1 mm crystal is below 3 THz. In optical rectification, the generated THz electric field is proportional to the source term $\omega I(\omega)$ where $\omega = 2\pi f$ is the angular frequency and $I(\omega)$ is the Fourier transform of the input optical pump pulse [29, 30]. Assuming frequency-independent phase matching, this term defines the spectral contents of the generated THz pulse. For example a pulse with 165 fs (FWHM) will match a Gaussian-like pulse with spectral half maxima at 0.73 THz and 4.37 THz. The short Ti:Sa pump pulse (45 fs) used in our experiment results in a much wider THz spectrum that is mostly lost due to phase mismatch between the group and phase velocity between the pump and THz radiation, respectively. We verified this claim experimentally by varying the pump pulse duration. To perform this in our laser system, we changed the spectral peak to 820 nm and then tuned the bandwidth. Preserving the transform limit, this corresponds to a variation in the pulse width from 34 to 74 fs (FWHM) under the approximation of a Gaussian-shaped profile. As we increased the pulse duration, we observed corresponding increase in the generated relative THz energy (keeping the total pulse energy constant) reaching up to 200 % as shown in Fig. 3a.

The increase rate was steep at the beginning, then slowed down. However, as we vary the pulse width (fixing the total energy and transform limit), the generation efficiency complicates with the nonlinear absorption in the crystal and the phase matching function. In-depth study of the parametric-dependence of the generation efficiency requires more tunability of the spectral and temporal properties of the pump pulse (as a function of the crystal thickness). However, we briefly studied the spectral dependence of the generation efficiency by inserting a slit in the grating pair compressor (thus fixing the pump bandwidth) and tuning it around 800 nm (Fig. 3b). We did so by placing a slit inside the compressor. We observed monotonic increase in the generation efficiency with the increase of the center wavelength nearly reaching a factor of 5 as we tuned from 782 nm to 822 nm (Fig. 3c).

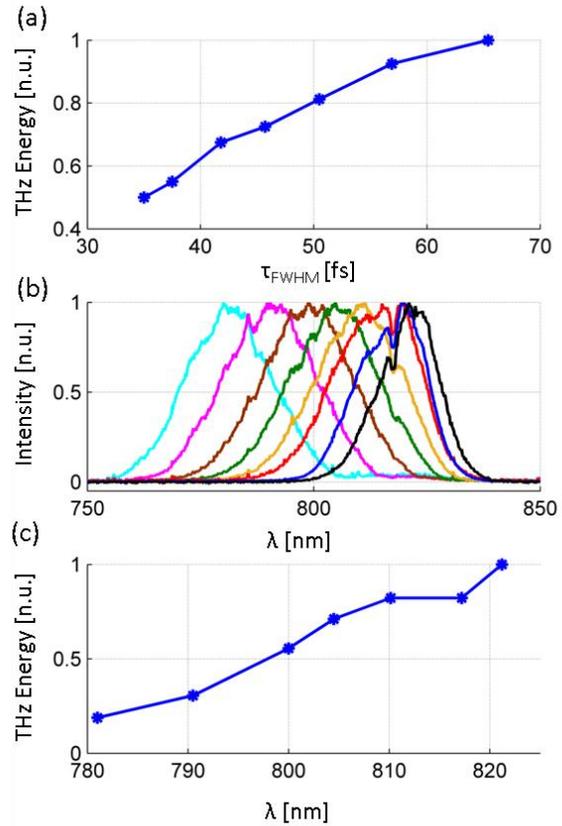

**Fig. 3.** Dependence of the generated THz pulse energy on the pump pulse temporal and spectral parameters. (a) The dependence of the THz output energy on the pump pulse duration. (b) Varying the pump spectrum center and (c) the corresponding change in the generation efficiency.

In conclusion, we have introduced the organic BNA crystal as an optical rectifier for intense single-cycle THz pulse generation using Ti:Sa pump technology. The high conversion efficiency (0.25%) and the low-frequency THz output (0.2-3 THz) makes this crystal a valuable alternative to LN sources commonly used in this frequency range. The simple generation scheme with a collimated pump beam offers a THz beam with excellent beam propagation properties which facilitate beam focusing and thus make the source attractive for THz strong-field applications. Our studies suggest that there is more space for efficiency optimization by controlling the pump pulse duration and spectrum. Future optimization of the BNA-based THz source is foreseen to further enhance the source performance for nonlinear THz spectroscopy applications.

**Acknowledgment:** We are grateful to Marta Divall, Alexandre Trisorio, and Andreas Dax from the Swiss FEL laser group for supporting the operation of the Ti:Sapphire laser system. We acknowledge the support from Edwin Divall in DAQ. We acknowledge financial support from the Swiss National Science Foundation (SNSF) (Grant No. 200021_146769). MS acknowledges partial funding from the European Community's Seventh Framework Programme (FP7/2007-2013) under grant agreement no. 290605 (PSI-FELLOW/COFUND). CPH acknowledges association to NCCR-MUST. S.B and K.T would like to acknowledge the Department of Science and Technology of the Government of India for the financial support vide grant No. SR/S2/CMP-0028/2011 Dated 19-01-2012.

**References**
1. S. Fleischer, Y. Zhou, R. W. Field, and K. A. Nelson, Phys. Rev. Lett. **107**, 163603 (2011).
2. M. Shalaby, C. Vicario, and C.P. Hauri, Nat. Commun. **6**, 8439 (2015).
3. A. T. Tarekegne, K. Iwaszczuk, M. Zalkovskij, A. Strikwerda, and P. U. Jepsen, New. J. Phys. **17**, 043002 (2015).
4. T. Kampfrath, K. Tanaka, and K.A. Nelson, Nat. Photon. **7**, 680 (2013).
5. O. Schubert, M. Hohenleutner, F. Langer, B. Urbanek, C. Lange, Y. Huttner, D. Golde, T. Meier, M. Kira, S. W. Koch, and R. Huber, Nat. Photon. **8**, 119 (2014).
6. B. Zaks, R.B. Lie, and M. S. Sherwin, Nature **483**, 580 (2012).
7. M. Shalaby and C. P. Hauri, Appl. Phys. Lett. **106**, 181108 (2015).
8. C. Vicario, C. Ruchert, F. Ardana-Lamas, P. M. Derlet, B. Tudu, J. Luning and C. P. Hauri, Nature Photon. (2013).
9. M. Shalaby, C. Vicario, and C. P. Hauri, New. J. Phys. **18**, 013019 (2015).
10. H. Hirori, A. Doi, F. Blanchard, and K. Tanaka, Appl. Phys. Lett. **98**, 091106 (2011).
11. M. Shalaby and C.P. Hauri, Nature Communications **6**, 5976 (2015).
12. C. Vicario, A.V. Ovchinnikov, S.I. Ashitkov, M.B. Agranat, V.E. Fortov and C.P. Hauri, Opt. Lett. **39**, 6632 (2014).
13. C.P. Hauri, C. Ruchert, C. Vicario, F. Ardana, Appl. Phys. Lett. 99, 161116 (2011).
14. C. Vicario, B. Monoszlai, and C. P. Hauri, Phys. Rev. Lett. **112**, 213901 (2014).
15. F. Blanchard, X. Ropagnol, H. Hafez, H. Razavipour, M. Bolduc, R. Morandotti, T. Ozaki, and D. G. Cooke, Opt. Lett. **39**, 4333 (2014).
16. C. Vicario, B. Monoszlai, Cs. Lombosi, A. Mareczko, A. Courjaud, J. A. Fülöp, and C. P. Hauri, Opt. Lett. **38**, 5373 (2013).
17. Shu-Wei Huang, Eduardo Granados, Wenqian Ronny Huang, Kyung-Han Hong, Luis E. Zapata, and Franz X. Kärtner, Opt. Lett. **38**, 796-798 (2013).
18. J. A. Fülöp, Z. Ollmann, Cs. Lombosi, C. Skrobol, S. Klingebiel, L. Pálfalvi, F. Krausz, S. Karsch, and J. Hebling, Opt. Lett. **22**, 20155 (2014).
19. B. Monoszlai, C. Vicario, and C. P. Hauri, Opt. Lett. **38**, 5106 (2013).
20. C. Somma G. Folpini, J. Gupta, K. Reimann, M. Woerner and T. Elsaesser, Opt. Lett. **40**, 3404, (2015).
21. P. Juranic et al., Opt. Express **22**, 30004 (2014).
22. E. A. Nanni, W. R. Huang, K. H. Hong, K. Ravi, A. Fallahi, G. Moriena, R. J. D. Miller, F. X. Kaertner, Nat.Commun. **6**, 8486 (2015).
23. H. Hashimoto, Y. Okada, H. Fujimura, M. Morioka, O. Sugihara, N. Okamoto, and R. Matsushima, Jpn. J. Appl. Phys. **36**, 6754 (1997).
24. K. Miyamoto, H. Minamide, M. Fujiwara, H. Hashimoto, and H. Ito, Opt. Lett. **33**, 252 (2008).
25. K.Kuroyanagi, M. Fujiwara, H. Hashimoto, H. Takahashi, S.Aoshima and Y. Tsuchiya, Jpn. J. Appl. Phys. **45**, 4068 (2006).
26. M. Fujiwara, M. Maruyama, M. Sugisaki, H. Takahashi, S. I. Aoshima, R. J. Cogdell, and H. Hashimoto, Jpn. J. Appl. Phys. **46,** 1528 (2007).
27. H. Hashimoto, H. Takahashi, T. Yamada, K. Kuroyanagi and T. Kobayashi, J. Phys.: Condens. Matter **13**, L529 (2001).
28. M. Shalaby, F. Vidal, M. Peccianti, R. Morandotti, F. Enderli, T. Feurer, and B. Patterson, Phys. Rev. B **88,** 140301 (R) (2013).
29. A. Schneider, M. Neis, M. Stillhart, B. Ruiz, R. U. A. Khan, and P. Günter, Opt. Lett. **23,** 1822 (2006).
30. M. Shalaby and C.P. Hauri, Sci. Reports **5**, 8059 (2015).